\documentclass[floatfix,twocolumn,amsmath,amssymb,aps,prb,showpacs,letter,superscriptaddress]{revtex4}
\usepackage{latexsym,epsfig,bm,times,psfrag,subfigure}
\usepackage{graphicx}
\usepackage{bm}
\usepackage{amssymb}
\usepackage{amsmath}
\usepackage{dcolumn}
\usepackage{color}
\usepackage{epstopdf}
\usepackage{float}

\newcommand{\szio}{Sr$_{3}$ZnIrO$_{6}$}
\newcommand{\szro}{Sr$_{3}$ZnRhO$_{6}$}
\newcommand{\smio}{Sr$_{3}$MgIrO$_{6}$}

\newcommand{\czmo}{Ca$_{3}$ZnMnO$_{6}$}

\newcommand{\cco}{Ca$_{3}$Co$_2$O$_{6}$}

\newcommand{\ccro}{Ca$_{3}$CoRhO$_{6}$}

\newcommand{\snio}{Sr$_{3}$NiIrO$_{6}$}

\newcommand{\cnmo}{Ca$_{3}$NiMnO$_{6}$}

\begin{document}
\title{Non-collinear Order and Spin-Orbit Coupling in Sr$_{3}$ZnIrO$_{6}$}
\author{Paul A. McClarty}
\altaffiliation{paul.mcclarty@stfc.ac.uk}
\affiliation{ISIS Neutron and Muon Source, STFC-Rutherford-Appleton Laboratory, Harwell Campus, Oxfordshire, OX11 0QX, UK}
\affiliation{Max Planck Institute for the Physics of Complex Systems, N\"{o}thnitzer Str. 38, 01187 Dresden}
\author{Adrian  D. Hillier}
\affiliation{ISIS Neutron and Muon Source, STFC-Rutherford-Appleton Laboratory, Harwell Campus, Oxfordshire, OX11 0QX, UK}
\author{Devashibhai T. Adroja }
\altaffiliation{devashibhai.adroja@stfc.ac.uk}
\affiliation{ISIS Neutron and Muon Source, STFC-Rutherford-Appleton Laboratory, Harwell Campus, Oxfordshire, OX11 0QX, UK}
\affiliation{Highly Correlated Electron Group, Physics Department, University of Johannesburg, P.O. Box 524, Auckland Park 2006, South Africa}
\author{Dmitry D. Khalyavin}
\affiliation{ISIS Neutron and Muon Source, STFC-Rutherford-Appleton Laboratory, Harwell Campus, Oxfordshire, OX11 0QX, UK}
\author{Sudhindra Rayaprol}
\affiliation{UGC-DAE CSR, Mumbai Center, R-5 Shed, BARC, Trombay, Mumbai 400085, India}
\author{Pascal Manuel}
\affiliation{ISIS Neutron and Muon Source, STFC-Rutherford-Appleton Laboratory, Harwell Campus, Oxfordshire, OX11 0QX, UK}
\author{Winfried Kockelmann}
\affiliation{ISIS Neutron and Muon Source, STFC-Rutherford-Appleton Laboratory, Harwell Campus, Oxfordshire, OX11 0QX, UK}
\author{Echur V. Sampathkumaran}
\affiliation{UGC-DAE CSR, Mumbai Center, R-5 Shed, BARC, Trombay, Mumbai 400085, India}
\affiliation{Tata Institute of Fundamental Research, Homi Bhabha Road, Colaba, Mumbai 400005, India}
\date{\today} 
\begin{abstract}
Sr$_{3}$ZnIrO$_{6}$ is an effective spin one-half Mott insulating iridate belonging to a family of magnets which includes a number of quasi-one dimensional systems as well as materials exhibiting three dimensional order. Here we present the results of an extensive investigation into the magnetism including heat capacity, a.c. susceptibility, muon spin rotation ($\mu$SR), neutron diffraction and inelastic neutron scattering on the same sample. It is established that the material exhibits a transition at about $17$ K into a three-dimensional antiferromagnetic structure with propagation vector $\boldsymbol{k}=(0,\frac{1}{2},1)$ in the hexagonal setting of R$\bar{3}$c and non-collinear moments of 0.87$\mu_{\rm B}$ on Ir$^{4+}$ ions. Further we have observed a well defined powder averaged spin wave spectrum with zone boundary energy of $\sim 5$ meV at $5$ K. We stress that a theoretical analysis shows that the observed non-collinear magnetic structure arises from anisotropic inter- and intra- chain exchange which has its origin in significant spin-orbit coupling. The model can satisfactorily explain the observed magnetic structure and spin wave excitations.
\end{abstract}


\maketitle
\section{Introduction} 
In materials with Ir$^{4+}$ ($5d^5$, $S=1/2$) ions occurring in IrO$_6$ octahedral units, spin-orbit entanglement combined with the crystal field lead to an effective description in terms of interacting spin one-half moments \cite{SrIrO}. Exploring such magnetic systems has been a topic of intense study over the last few years motivated by the search for novel electronic phases including topological insulators, complex magnetic structures, quantum spin liquids and other exotica. For example, in various 2D and 3D honeycomb lattice magnets, strong spin-orbit coupling leads to Ising exchange along bonds joining magnetic ions which may lead to Kitaev spin liquids being realized in solid state magnets  \cite{Jackeli2009,Chaloupka2010,Choi2012,Chun2015}. The iridate pyrochlores have provided a rich set of interesting phenomena and inspired new ideas to realize novel phases \cite{Review}. More recently, the electron doped layered iridate Sr$_2$IrO$_4$ has been shown to have tantalizing similarities to the hole-doped precursors of the high temperature superconductors \cite{SrIrO}. 

In this paper, we consider iridate magnetism in yet another context: in \szio\ - one of a family of magnetic materials with chemical formula A$_3$BB'O$_6$ in which magnetic ions live on the B or B' positions or both and where A is either Sr$^{2+}$ or Ca$^{2+}$. Recent interest in this family of materials among condensed matter physicists stems from the fact that when both B and B' are magnetic the materials often exhibit quasi-1D Ising magnetism and that the magnetic chains have a triangular coordination giving the potential for geometrical frustration to affect their ground states \cite{GenRef1,GenRef2,Lefrancois2014,Ou2014}. The most well-known case is \cco\ \cite{CCO,CCO2,CCO3,CCO4, GenRef1,GenRef2} which has ferromagnetically coupled chains with weak antiferromagnetic inter-chain correlations.  
 However, many of these materials are phenomenologically interesting: there is evidence for partially disordered antiferromagnetism at low temperatures in  \cco\ \cite{CCO,CCO2,CCO3,CCO4}, \ccro\ \cite{CRO} and \snio\cite{Mikhailova2012} co-existing with peculiar dynamical properties including memory effects. In sharp contrast, \cnmo\ \cite{Kawasaki} exhibits spiral order at low temperatures while Ca$_3$CoMnO$_6$ exhibits collinear-magnetism-driven ferroelectricity \cite{Choi2008}. In all these materials, the inter-chain coupling plays a crucial role. In order to appreciate better the nature of these couplings, one may substitute nonmagnetic ions onto the B site so that the inter and intra chain couplings become comparable which is the case in \szio\ \cite{Nguyen1995}. 
 
Early work on \szio\  reported the crystal structure determined from x-ray measurements and both the moment size and the scale of the exchange coupling from the d.c. susceptibility but no 3D long-range magnetic ordering transition was observed in the neutron diffraction work \cite{Nguyen1995,LampeOnnerud96}.  Subsequent studies based on a.c. and d.c. magnetization as well as heat capacity ($C_p$) measurements established the existence of long-range magnetic ordering in this material \cite{Niazi200x}. In this paper, we present results of an extensive experimental investigation into the magnetic properties of this material including a.c. susceptibility, heat capacity, muon spin rotation ($\mu$SR), neutron diffraction and inelastic neutron scattering culminating in the determination of the magnetic structure in this material and its excitations. The magnetism of \szio\ is {\it a priori} expected to be affected significantly by spin-orbit coupling.  We present a theoretical model that explains the non-collinear spin structure in this material based on exchange parameters which rely on the presence of significant spin-orbit coupling. Our theory establishes the most general form of the couplings that can arise between magnetic B site ions.

\begin{figure}
\centering
\includegraphics[width=\columnwidth, trim={0cm 4cm 4cm 0cm},clip]{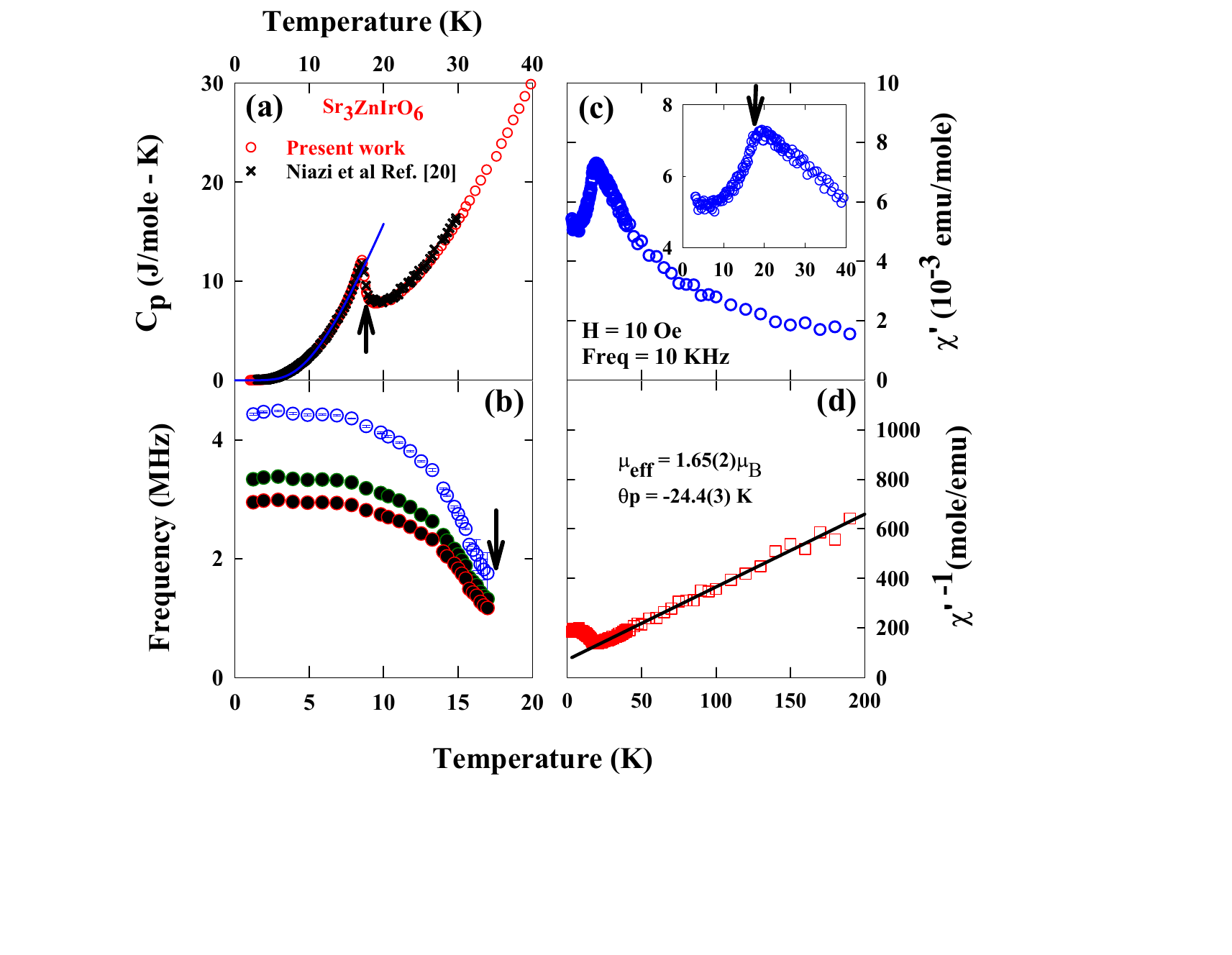}
\caption{ 
{(Color online)}
(a) Heat capacity of \szio\  and for comparison we have also shown the heat capacity data from the published work of Niazi et al, Ref. [21]. The solid line shows a fit (see  text). (b) Triplet of $\mu$SR frequencies vs. temperature showing onset at $T_{\rm N}$ marked by an arrow. (c) ac-susceptibility as a function of temperature with inset showing vicinity of  $T_{\rm N}$. (d) Inverse ac-susceptibility as a function of temperature. The solid line shows a fit to Curie$-$Weiss law in the paramagnetic regime.
\label{fig:1}}
\end{figure}

\section{Experimental Methods} 
The compound \szio\ was prepared by the solid-state route as detailed in Ref.~\onlinecite{Niazi200x}. The quality of the sample was checked using x-ray and neutron diffraction measurements and found to be a single phase. The ac-magnetization and heat capacity, $C_p$, measurements were carried out using a Quantum Design physical properties measurement system  (PPMS). The neutron diffraction measurements were carried out on the GEM and WISH time-of-flight diffractometers at the ISIS facility. The powder sample of \szio, contained in a vanadium can of $3$-mm diameter, was placed inside a standard helium cryostat under He-exchange gas on WISH. The temperature dependent order parameter was measured on the GEM diffractometer. The $\mu$SR experiment was performed on the MuSR spectrometer at the ISIS facility.  The inelastic neutron scattering measurements were carried out on the high flux time of flight spectrometer MERLIN at the ISIS Facility using an incident neutron energy E$_{i}$= 18 meV at 5~K and 25~K. The energy resolution at the elastic position was ~1 meV (FWHM). The powder sample of Sr$_3$ZnIrO$_6$ (weight 4.1g) was mounted in an aluminium can with 40 mm diameter in an annular geometry to reduce neutron absorption.
 
\section{Experimental Results and Discussion} 

\subsection{Susceptibility and Heat capacity}

Figure~\ref{fig:1} shows the temperature dependent heat capacity, a.c. magnetic susceptibility and $\mu$SR frequency. As one can see in Fig.~\ref{fig:1}(a) in zero field a clear lambda-type anomaly is observed near $17$ K in $C_p(T)$, which is indicative of a transition to long-range magnetic order. Further, the values of the heat capacity at the $\lambda$-anomaly of our data agree very well with the published data \cite{Niazi200x}, (Fig.~\ref{fig:1}(a)) indicating that the quality of the samples is similar. To estimate the gap in the spin wave spectrum we have fitted  the heat capacity data of Sr$_3$ZnIrO$_6$ between $2$ K and $16$ K using the formula
$C_p (T)= \beta T^3 +a \exp(-\Delta /T)$
where $\beta$ is related to Debye temperature,  $\Theta_{D}={(12{\pi}^{4}Rn/5\beta)}^{1/3}$, R is the molar gas constant and n is the number of atoms per formula unit. The parameter $\Delta$ in the last term is the magnetic spectral gap. We have fitted the  $C_p (T)$ between 2K and 16K to this formula. The value of the parameters obtained from the fit are  $\beta$=3.48(0.5)x10$^{-4}$(J/mole-K$^{4}$), $a$=89.5(3.7) and $\Delta$=38.57(0.6) K (3.32 meV). The estimated value of the spin gap is in agreement with that seen in the inelastic neutron scattering study $\sim$ 4 meV (see Fig. 3 and also Fig. 10 in the supplementary material (SM) \cite {SM}). From the value of $\beta$ we estimated $\Theta_{D}$=394.4~K.

The ac susceptibility was measured between $2$ and $300$ K in a magnetic field of $10$ Oe over a range of frequencies. The real part of the ac susceptibility $\chi'$ shows a peak at  $17$ K due to an antiferromagnetic ordering of the Ir moments (Fig.~\ref{fig:1}(c)) which is independent of frequency between $100$ Hz and $10^4$ Hz (see Fig.6(b) in SM \cite{SM}).  For temperatures greater than $25$ K the inverse susceptibility vs. temperature plot follows the expected Curie-Weiss behavior as shown in Fig.~\ref{fig:1}(d), which gives an effective magnetic moment, $1.65(2)$$\mu_B$ and the paramagnetic Curie-Weiss temperature,  $-24.4(3)$ K. The value of the observed magnetic moment is close to the $1.73$ $\mu_B$ expected for $S=1/2$ in agreement with the previous studies \cite{LampeOnnerud96,Niazi200x}. 

\subsection{Muon spin rotation and relaxation} 
The temperature dependence of the muon spin relaxation was measured in a zero field. For all temperatures above $T_{\rm N} =17$ K, the muon spin relaxation spectra can be described by a simple exponential decay. However, on cooling below $17$ K, coherent oscillations are observed in the $\mu$SR spectra (see Fig.~7 in SM \cite{SM}). This shows that we have a long-range magnetically ordered ground state. The $\mu$SR spectra are well described by three sinusoidal functions with a Gaussian envelope and an exponential decay. The temperature dependence of the three observed $\mu$SR frequencies is plotted in Fig.~\ref{fig:1}(b).  We have seen three well resolved frequencies up to $T_{\rm N}$. The overall temperature dependence of the observed frequencies is consistent with that expected for a second order phase transition. Irrespective of the absolute positions of the muons, the number of frequencies observed is determined by the local symmetry of the Ir$^{4+}$ moment. In the presence of threefold symmetry (i.e., Ir$^{4+}$ moment along the c axis) only one frequency will be observed. However, if the threefold symmetry is lost, which is the case for Sr$_{3}$ZnIrO$_{6}$ below the magnetic ordering transition, and the magnetic moment has a component perpendicular to the chains then we can see by symmetry that the muons will experience three different magnetic fields \cite{Hillier2011}.

\subsection{Neutron diffraction}
\begin{figure}
\centering
\hspace{3cm}
\centering
\includegraphics[width=9cm]{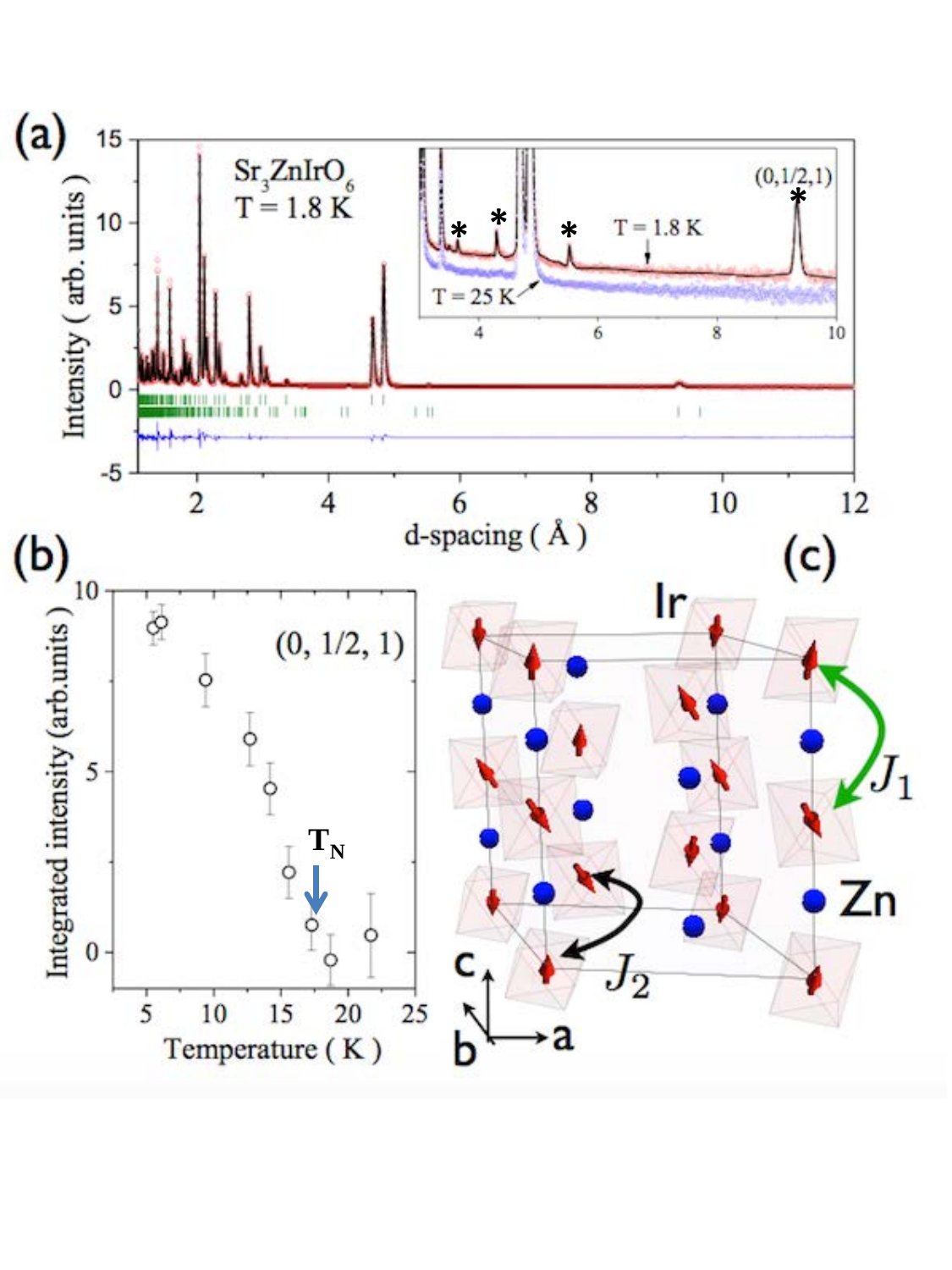}
\caption{ 
{(Color online)}
(a) Neutron diffraction pattern at 1.8K of \szio\ from the bank-$2$ at $58.2$ degrees of WISH. The symbols show the experimental data and the (black) line shows the fit using nuclear and magnetic structures. The (blue) line at the bottom shows the difference plot and vertical bars (green) indicate the Bragg peak positions: nuclear at the top and magnetic at the bottom. The inset shows the data on an expanded scale for 1.8 K (red color symbols) and 25 K (blue color symbols) where stars show the observed magnetic Bragg peaks. The solid line shows the fit to 1.8 K data including nuclear and magnetic structures. (b) The temperature dependent intensity of magnetic (0, 1/2, 1) peak measured on GEM. (c) The crystal structure in the hexagonal setting with Ir (red), Zn (blue) and the oxygen octahedra.  The magnetic structure in the cell is also shown together with the nearest neighbor and next nearest neighbor Ir-Ir distances as indicated by $J_1$ (green) and $J_2$ (black) respectively. 
\label{fig:2}}
\end{figure}

Figure~\ref{fig:2} shows a neutron diffraction pattern collected at $1.5$ K from the WISH diffractometer. At this temperature, there are four weak additional Bragg peaks (marked with stars) compared to the pattern collected in the paramagnetic state at $25$~K (see inset Fig.~\ref{fig:2}). The fact that these observed magnetic peaks disappear above $17$ K (see inset Fig.~\ref{fig:2}), indicates that the magnetic ordering temperature is $17$~K in agreement with the $T_{\rm N}$ determined by the $\mu$SR and heat capacity measurements. The intensities of the superstructure peaks decrease with momentum transfer ({\it Q}) confirming their magnetic origin. The peak positions can all be indexed by the propagation vector $\boldsymbol{k} = (0,1/2 ,1)$ in the hexagonal setting of the R$\bar{3}$c space group. This is a high-symmetry point of the Brillouin zone, labeled k$_5$ in the Kovalev notation and equivalent to $\boldsymbol{k} = (1/2 ,1/2 ,0)$ in the rhombohedral setting. The symmetry analysis shows that there are four one-dimensional representations in the little group, labelled $\tau_i$ ($i = 1, 4$) in Kovalev notation, \cite{Kovalev} but only two enter in the decomposition of $\Gamma = 3\tau_1 + 3\tau_3$. Both IR have three basis vectors, $\phi_{1, i}$ and $\phi_{3, i}$ ($i=1,2,3$). We first tried to fit the magnetic structure using $\tau_3$, but we were unable to fit the intensities and positions of all observed magnetic reflections  (see  Fig.~8(a) in the supplementary section).  Considering  $\tau_1$, none of the $\phi_{1, i}$ modes taken separately give a satisfactory agreement with the data. However, a solution is found by refining a model with a linear combination of the three modes. The final refinement yields a good description of the experimental data (Fig.~\ref{fig:2}(a)) and a magnetic Bragg factor R$_{\rm mag}$ = 14\% obtained using 
the software FULLPROF \cite{FULLPROF}. The mixing coefficients for the three modes $\phi_{1, i}$, $i = 1,2, 3$ are respectively $-0.072$, $-0.579$   and $0.670$ and the ordered state Ir moment of $0.87(3)$ $\mu_{\rm B}$. It is to be noted that the moment has three components  $-0.072$,  $-0.579$  and  $0.670$ along the $a$, $b$- and $c$-axis, respectively. Thus the magnetic structure is non-collinear, with a substantial angular deviation of the magnetic moment with respect to the $c$ axis. The collinear structures with moments either in plane or along the c-axis provide much worse fitting quality, yielding the magnetic Bragg factors 96\% and 27\%, respectively. In the proposed magnetic structure, the threefold symmetry axis is lost, and the magnetic transition corresponds to a change of magnetic point group from 3m.1' in the paramagnetic state to 2/m.1' in the magnetically ordered state. The observation of three frequencies in the muon data is also consistent with the loss of threefold symmetry that splits the oxygen sites into three inequivalent sites below the magnetic ordering as described above. A representation of the magnetic structure is displayed in Fig.~\ref{fig:2}c.

\begin{figure}
\includegraphics[width=8cm]{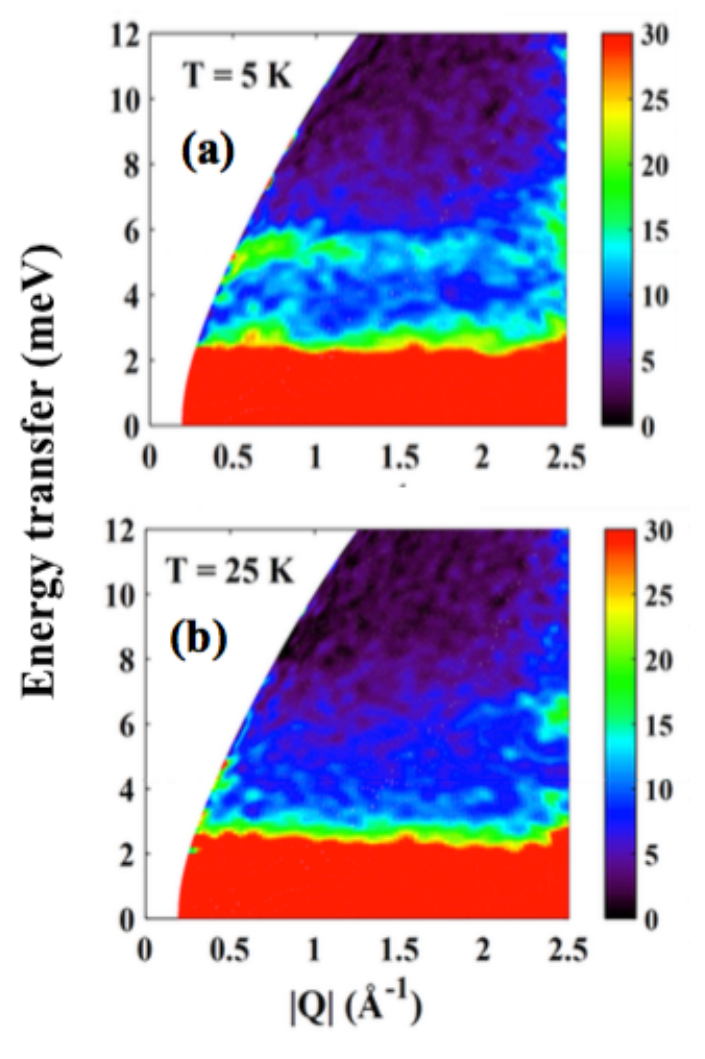}
\caption{ 
{(Color online)} Neutron scattering intensity maps.  Energy transfer vs. momentum transfer measured with an incident energy of $E_i=18$ meV on MERLIN for two different temperatures (a) one below and (b) one above $T_{\rm N}$.
\label{fig:3}}
\end{figure}
\subsection{Magnetic Excitations}
The inelastic neutron scattering (INS) intensity maps of Sr$_3$ZnIrO$_6$ measured at $5$~K and $25$~K with $E_i=18$ meV are shown in Fig.~\ref{fig:3}. At $T=5$ K, with the momentum transfer $\vert Q\vert$ below $1.5$ \AA$^{-1}$, a strong inelastic scattering peak which shows weak dispersion can be clearly observed between $3$ and $6$ meV. The magnetic origin of the strong scattering in the $3-6$ meV band is established by the initial reduction in the energy-integrated inelastic intensity as a function of $\vert Q\vert$. Further, at $25$ K there is no sign of scattering between  $3-6$ meV. We attribute the observed magnetic excitations between $3$ and $6$ meV for $\vert Q\vert$ below $2$ \AA$^{-1}$ to spin wave scattering from Ir$^{4+}$ ions in the magnetically ordered state below $17$ K , as expected. The spin wave excitations do not survive above $T_{\rm N}=17$ K, which is consistent with the proposed similarity between inter and intra chain interactions. In contrast, the spin wave excitations in Sr$_3$NiIrO$_6$ survive up to $250$ K, which is well above $T_{\rm N}=70$ K in that material, indicating the existence of a one-dimensional magnetic exchange interaction \cite{Toth2016}.

\section{Theoretical model for Exchange Couplings} 
The existence of three dimensional magnetic order without a regime of purely 1D correlations above $T_{\rm N}$ indicates that the scale of the further neighbor exchange is similar to that of the intra-chain exchange in \szio. In an iridate one expects, {\it a priori}, the spin-orbit coupling to be relevant to the low temperature magnetic order \cite{Yamaj2014}. This expectation is confirmed by the discrete nature of the symmetry breaking into a state with moments in the $ab$ plane that are orientated almost along the bond directions. In order to understand the origin of the magnetic structure of \szio, isotropic exchange is therefore not sufficient so we determine the most general bilinear couplings allowed by the lattice symmetries and then establish the minimal set of couplings that are necessary to select the observed magnetic structure.

For nearest neighbors within the chains along the $c$ axis, there are three possible types of exchange (see SM for detail \cite{SM}): Ising, isotropic easy plane and Dzyaloshinskii-Moriya \cite{DM}. The latter coupling has $\boldsymbol{D}$ vectors along the $c$ direction and alternating in sign from one bond to the next. The second neighbor interactions couple iridium ions in different chains. The chains are arranged in a triangular lattice but with three different staggerings in the $c$ direction such that the second neighbor coordination is in the form of a puckered hexagon shown in Fig. 9 of the SM \cite{SM}. One may show that there are six distinct types of second neighbor exchange coupling. Of these six, one couples the Ising components and three couple the $XY$ components, one of which preserves a global $U(1)$ symmetry. The remaining two couple transverse to $Z$ components and are central to selecting the observed structure in \szio. Further details on the symmetry analysis and explicit forms for the second exchange can be found in the Supplementary Material. 

{\it Classical Phases} $-$ Having established the nature of the exchange couplings in \szio\ and related materials with significant spin-orbit coupling, we may find the possible magnetically ordered phases. We obtain the ground states of the exchange model by rotating the vector spins along their local mean field on a periodic cluster. As guidance for this calculation - in particular, to choose the correct minimal cell for the minimization - we obtain the ordering wavevector $\mathbf{Q}_{\rm ord}$ from the location $\mathbf{q}=\mathbf{Q}_{\rm ord}$ of the minimum eigenvalue of the Fourier transformed interactions ${\cal J}_{ab}(\mathbf{q})$ where $a$ and $b$ are the $6$ magnetic sublattices in the hexagonal representation of the interactions. 

Exchange to nearest neighbor only couples moments within chains running along the $c$ axis. Further neighbor couplings are necessary for fully three dimensional ordered structures. For example, the second neighbor XY antiferromagnetic coupling leads to spontaneous symmetry breaking to a collinear XY structure identical to the one observed in \czmo ~\cite{Ruan2014}. A collinear structure with moments along the $c$ axis is found for second neighbour Ising exchange. 

One obtains the structure close to that observed in \szio\  and \szro\ by including couplings for exchanges $H_{z\pm,{\rm as}}$ (antisymmetric) and $H_{{\rm z}\pm,{\rm s}}$ (symmetric) in the ratio $J^{(2)}_{{\rm z}\pm,{\rm s}}/J^{(2)}_{{\rm z}\pm,{\rm as}}=\sqrt{3}$ with both couplings negative. Then one may switch on either antiferromagnetic nearest neighbor Ising exchange $J^{(1)}_{\rm zz}$ or second neighbor ferromagnetic Ising exchange $J^{(2)}_{\rm zz}$. The phase diagram obtained with fixed $J^{(2)}_{{\rm z}\pm,{\rm s}}/J^{(2)}_{{\rm z}\pm,{\rm as}}$ and variable $J^{(1)}_{\rm zz}$ and $J^{(2)}_{\rm zz}$ is shown in Fig.4. There are two distinct ground states. The one for small $\vert J^{(2)}_{\rm zz}\vert$ (colored purple in Fig.4) is a simple collinear ferromagnet with $\mathbf{Q}_{\rm ord}=0$ and the other is proximate to the \szio\  and \szro\ \cite{Hillier2011} antiferromagnetic structure with $\mathbf{Q}_{\rm ord}=(0,1/2,1)$. The moments are canted away from the $c$ axis by an angle which is represented by the (colored) contours in Fig.4. The experimental value of about $39$ degrees is indicated by a black dashed line.

\begin{figure}[htpb]
\centering
\includegraphics[width=8cm]{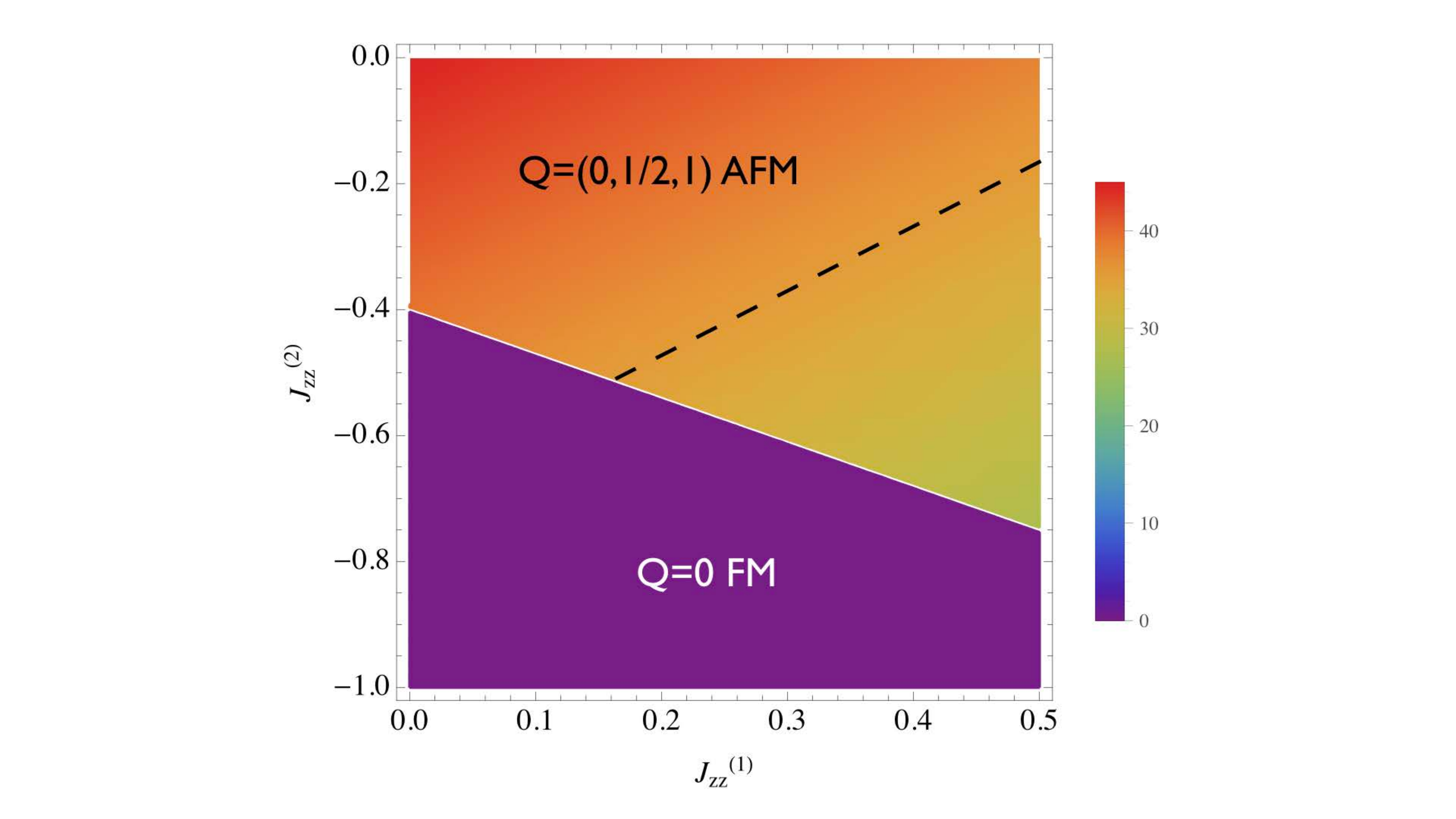}
\caption{ 
{(Color online) Phase diagram with two types of Ising exchange and $J^{(2)}_{z\pm,{\rm s}}$, $J^{(2)}_{z\pm,{\rm as}}$ fixed.}
The diagram shows two phases. At the bottom of the diagram, the blue region is the collinear ferromagnetic phase. The second phase is the magnetic structure of \szio\ up to a small rotation of the moments in the XY plane. The gradient of colors in the latter phase represents the canting angle of the moments with respect to the $c$ axis with the color bar in degrees. The dashed line indicates the range of parameters over which the canting angle of the moments from the c axis matches the experimental value for Sr$_3$ZnIrO$_6$ (see Fig.2(c)). 
\label{fig:PhaseDiagram}}
\end{figure}

The \szio\  and \szro\ structures differ from the $(0,1/2,1)$ phase shown in Fig.4 by a staggered rotation $\delta\phi_i$ of the moments in the $ab$ plane away from the $60n$ degree angles where the magnitude of $\delta\phi_i$ is the same for all sites. The relevant rotation angle is about $6$ degrees for  \szio. Just such a rotation can be achieved by switching on an antiferromagnet XY coupling between nearest neighbor moments.
\begin{figure}
\includegraphics[width=8cm]{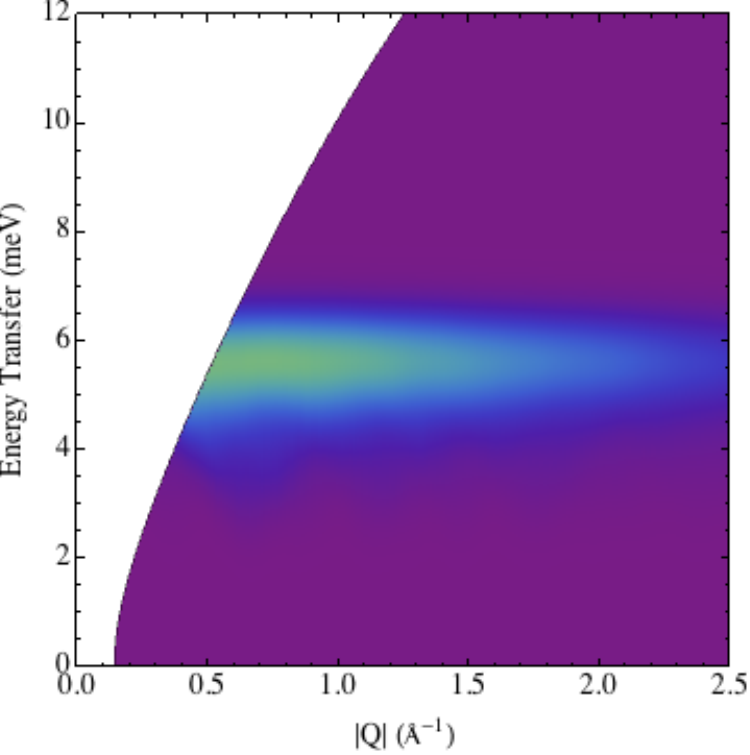}
\caption{ 
{(Color online)Angular averaged spin wave intensity computed to linear order in $1/S$}. The exchange parameters are only weakly constrained by the observed magnetic structure. For the spin wave calculation we chose the following couplings (in meV) $J^{(1)}_{\rm zz}=0.3$,  $J^{(1)}_{\rm xy}=0.135$ to first neighbor and $J^{(2)}_{\rm zz}=-0.525$, $J^{(2)}_{\pm}=0.263$, $J^{(2)}_{z\pm,{\rm as}}=-0.18$ and $J^{(2)}_{z\pm,{\rm s}}=-0.311$ to second neighbor.  
\label{fig:SW}}
\end{figure}

On the basis of the exchange model presented above, we compute the spin wave excitations to leading order in a Holstein-Primakoff $1/S$ expansion. The spectrum has $12$ modes albeit with some degeneracies. The spectrum is gapped as we should expect on the basis of the observed discrete symmetry breaking. We are able to obtain the absolute scale of the exchange parameters on the basis of a comparison between the experimental data and the calculated scattering. The powder averaged dynamical structure factor for a set of parameters that give the correct magnetic structure as their semiclassical ground state is shown in Fig.~5. A form factor for Ir$^{4+}$ has been applied using parameters from Ref.~\onlinecite{Kobayashi2011}. The model captures the principal features of the INS data including the single strong band of intensity between $4$ meV and $6$ meV which is clearly visible in both plots and a maximum between $0.5$ and $1$\AA$^{-1}$. While our exchange parameters are consistent with the available data there is some freedom to choose these parameters while holding the magnetic structure fixed. Single crystal inelastic neutron scattering may be used to determine these parameters completely.

\section{Conclusion}
We have studied the magnetism in \szio\ using a variety of techniques including heat capacity, ac susceptibility, $\mu$SR, neutron diffraction and inelastic neutron scattering. These reveal a clear magnetic transition at $17$ K into a fully 3D long range ordered magnetic structure with propagation vector $(0,1/2,1)$ in which the moments in the ordered phase exhibit a discrete coplanar orientation with respect to the principal lattice directions. We have also presented a theoretical analysis with which we have understood the origin of the observed magnetic structure.  By determining the complete set of allowed couplings between first and second neighbors we established the importance of anisotropic inter-chain exchange between the XY component of one moment and the Ising component of its neighbor. The model presented here is applicable to both the simple structure observed in Ca$_3$ZnMnO$_6$ and the complex non-collinear ordering reported here which is also present in \szro\ \cite{Hillier2011} and \smio\ 
\cite{Unpublished}.

\section*{Acknowledgements}

P.McC. acknowledges an STFC Keeley-Rutherford fellowship. D.T.A. and A.D.H. acknowledge financial assistance from CMPC-STFC grant number CMPC-09108. D.T.A. would like to thank JSPS for an invitation fellowship to visit Hiroshima University.  One of us (E.V.S) would like thank  SERB, Government of India, for J.C. Bose Fellowship.

\clearpage

\onecolumngrid
{\textnormal
{\bf {\Large{Supplementary Material: Non-collinear Order and Spin-Orbit Coupling in Sr$_{3}$ZnIrO$_{6}$}}}

\section*{}
{\bf {\Large {I. Magnetization and Susceptibility}}}
\\*
\begin{figure} [htpb]
\centering
\includegraphics[width=0.5\columnwidth]{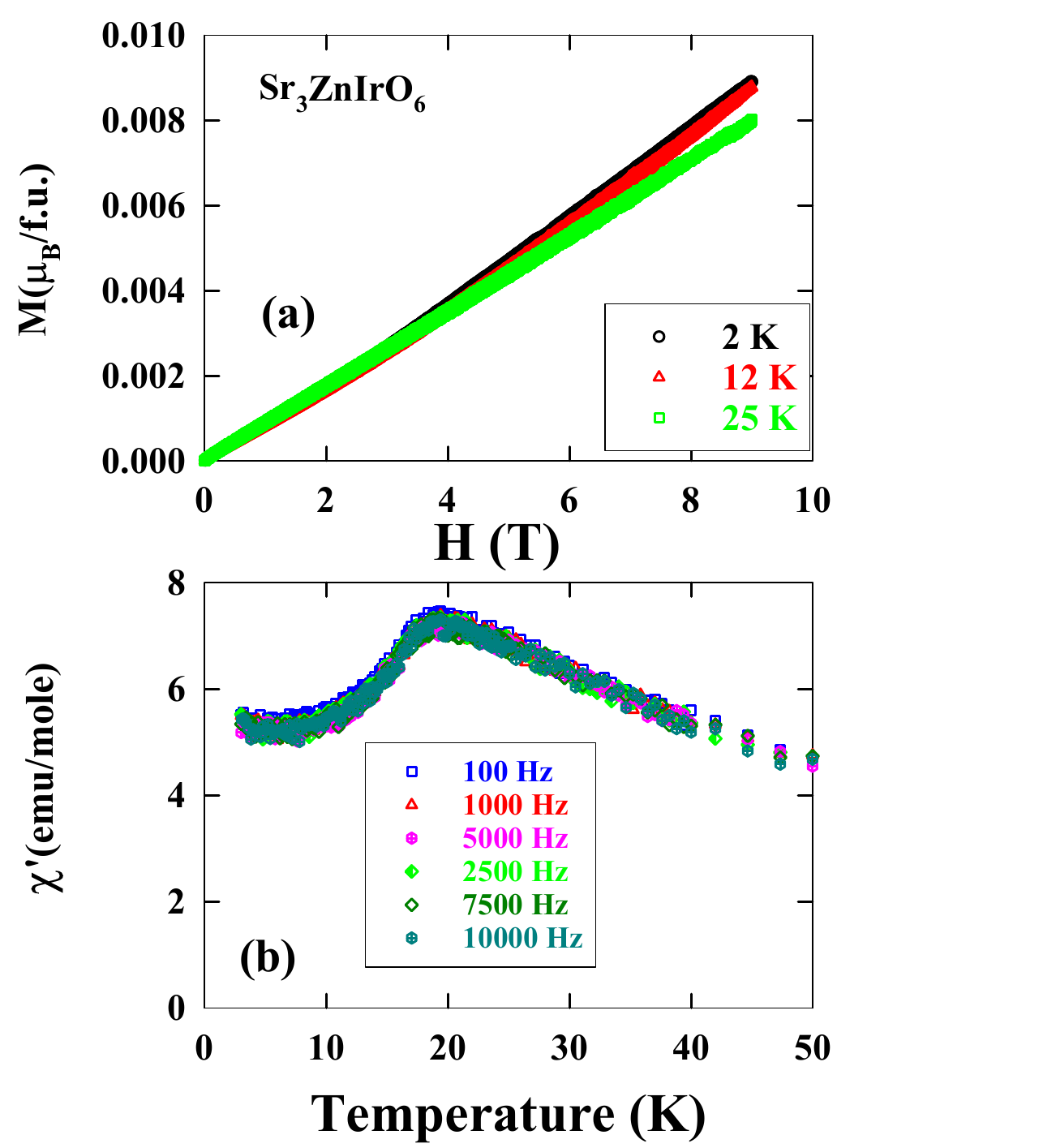}
\caption{
(Color online) (a) Magnetization isotherms at various temperatures of Sr$_{3}$ZnIrO$_{6}$. (b) Figure illustrating the frequency independence of the real part of the a.c. susceptibility in the vicinity of the magnetic phase transition.
\label{fig:CVM}}
\end{figure}

\clearpage
\section*{}
{\bf {\Large {II. $\mu$SR asymmetry spectra}}}
\\*
\begin{figure}[htpb]
\includegraphics[width=0.5\columnwidth]{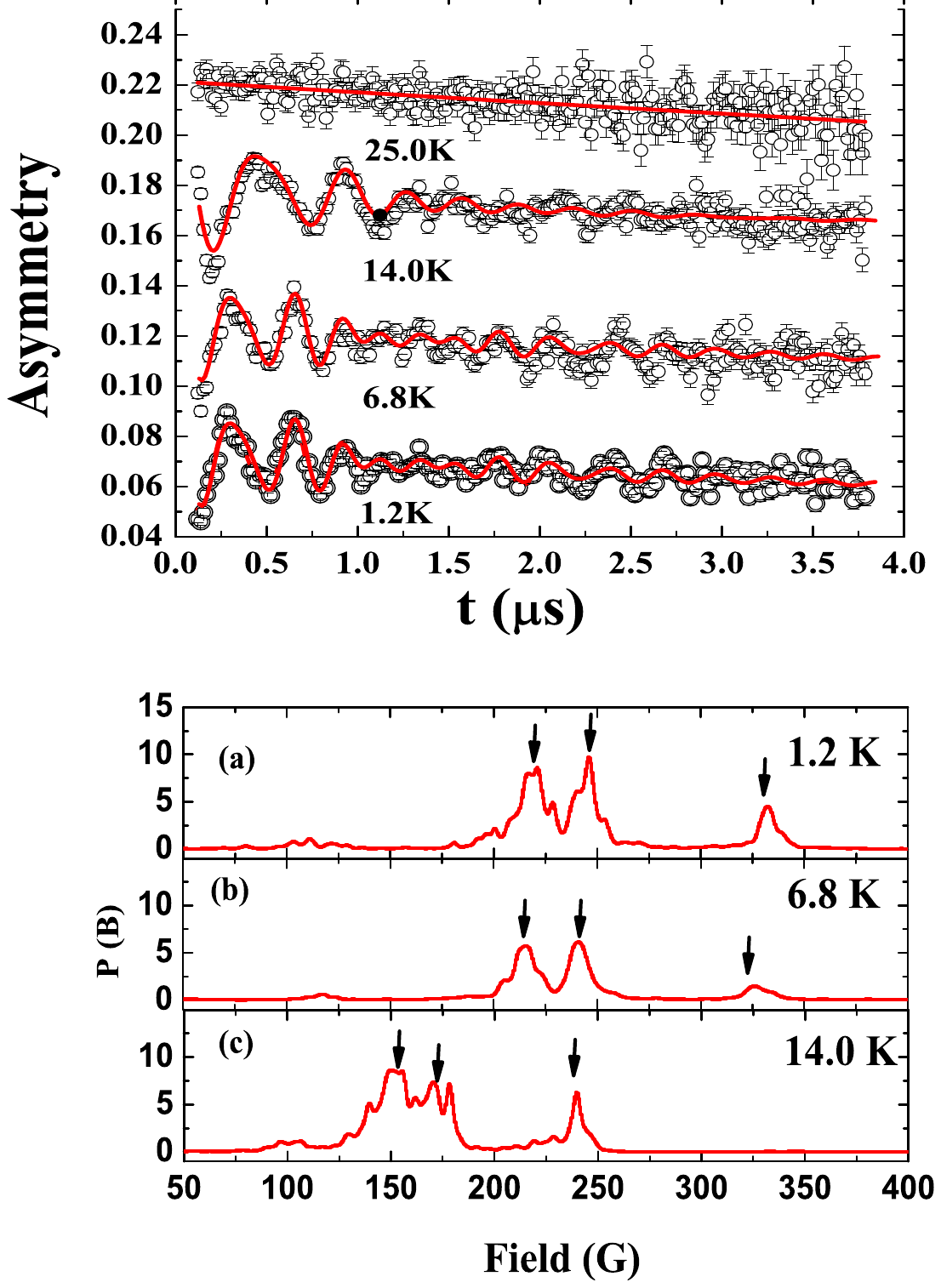}
\caption{ 
{ (Color online) (top) $\mu$SR asymmetry measured at four different temperatures. Oscillations are clearly visible at $14$ K and below. Analysis of this data leads to the three distinct frequencies plotted in Fig.~1 (b) of the main paper. (bottom) The field distribution (using a maximum entropy technique) from the muon spectra  given in the top figure. Three field values are corresponding to observed three frequencies in the $\mu$SR spectra.}  
\label{fig:asymmetry}}
\end{figure}
\clearpage
\section*{}
{\bf {\Large{III. Neutron diffraction analysis using the two different irreducible representations}}}
\\*
\begin{figure}[htpb]
\includegraphics[width=0.8\columnwidth, trim={0 0 0 0},clip]{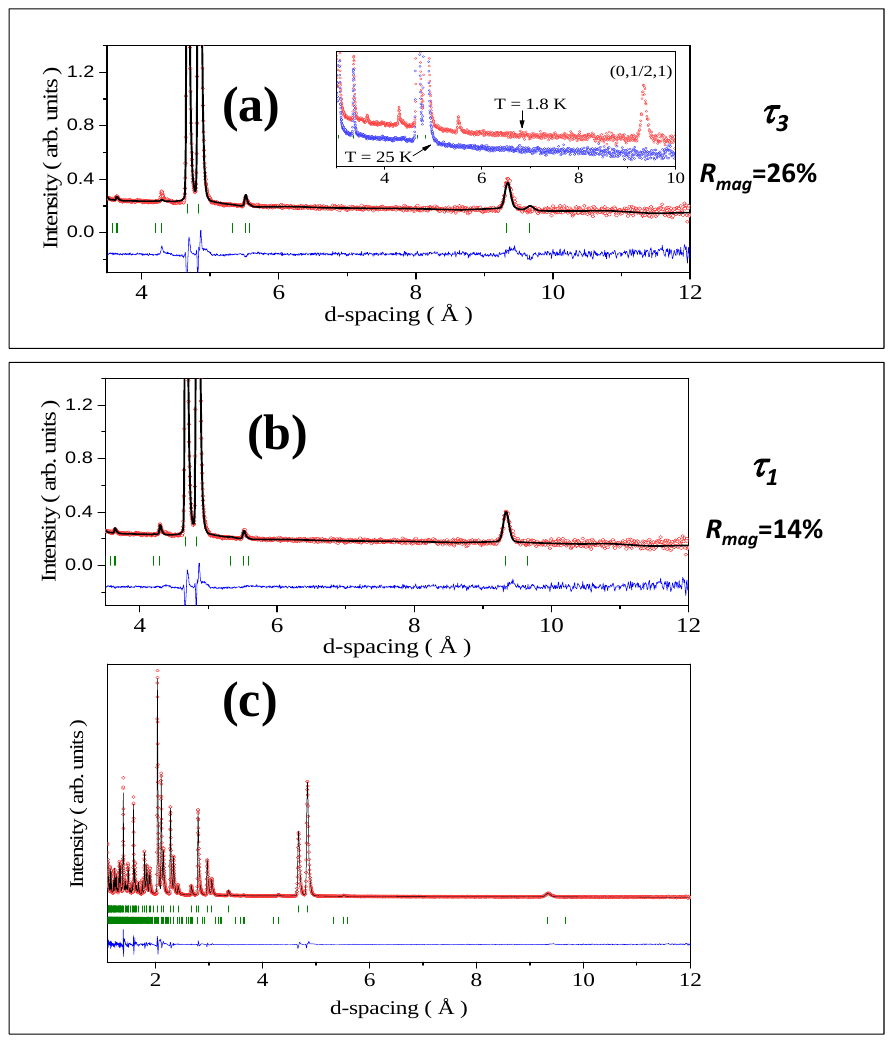}
 \caption{ 
{ (Color online)  Neutron diffraction pattern of Sr$_{3}$ZnIrO$_{6}$ collected at 1.8 K and fitted using the two different irreducible representations (a) $\tau$$_{3}$ and (b-c) $\tau$$_{1}$ to model the magnetic intensities. The symbols show the experimental data and the line shows the fit using nuclear and magnetic structures. The blue line at the bottom shows the difference plot and vertical bars indicate the Bragg peak positions: nuclear at the top and magnetic at the bottom. The insets show diffraction patterns collected at 1.8 K and 25 K at a large d-spacing range, where the magnetic peaks are observed.}  
\label{fig:diff}}

\end{figure}

\clearpage
\section*{}
{\bf {\Large {VI. Symmetries}}
\\*

{\textnormal
The starting point for our analysis is the set of crystal symmetries. Here we refer to the hexagonal setting of the R$\bar{3}$c space group. The point group symmetry at the magnetic sites (Wyckoff sites $6b$) is S$_6$ which, apart from the trivial element, is composed of $3$-fold rotations about the $c$ axis, inversion and S$_6$ operations which are $6$-fold rotations about the $c$ axis followed by reflection through the $ab$ plane passing through the origin. There are two other symmetry operations. The first is a reflection about one of the three planes containing a magnetic ion and two of its (hexagonally coordinated) second neighbors followed by a translation in the $c$ direction by $(0,0,1/2)$. The second is a rotation about one the three axes in the $ab$ plane passing through the magnetic ion and the midpoints of two of its second neighbors followed by a translation along the $c$ axis through $(0,0,1/2)$. 

There are two magnetically inequivalent sites. We label sites $i,a$ using the convenient body centered rhombohedral setting where $a=1,2$ is the sublattice and $i$ is the rhombohedral cell. The nearest neighbor couplings connect the two different sublattices. With this exchange alone, the magnetic structure splits up into decoupled chains. Each site has six second nearest neighbors which also connect different sublattices.}

\section*{}
{\bf {\Large {V. Exchange Couplings between $B$ ions}}}
\\*
{\textnormal We first consider the number of exchange parameters allowed between nearest neighbors. The $2$ nearest neighbors of each magnetic ion are in the $c$ direction and their coupling alone results in a lattice of decoupled chains. In this case we can write down the allowed couplings with little analysis. There are three such couplings: one is the Ising coupling $\sum_{i,a,b} S^z_{i,a} S^z_{i,b}$ where $i$ runs along the chains, the second is the isotropic XY coupling $(1/2) \sum_{i,a,b} S^+_{i,a} S^-_{i,b}+S^-_{i,a} S^+_{i,b}$. The third is the Dzyaloshinskii-Moriya interaction $\sum_{i,a,b} \mathbf{D}_{a \rightarrow b}\cdot(\mathbf{S}_{i,a} \times \mathbf{S}_{i,b})$ with $\mathbf{D}_{a\rightarrow b}$ vector along the $c$ axis and alternating in sign between adjacent bonds.

Now we focus on the couplings between second neighbors. First we can compute the number of second neighbor exchange parameters allowed by the local point group symmetry. Then we impose the remaining constraints coming from the symmetries that involve translations. 

The second neighbors form a puckered hexagon around the central magnetic ion (see Fig.~\ref{fig:Coordination}). Before imposing symmetry constraints, each bond contributes $9$ couplings of the form $S_i^{\alpha}S_j^{\beta}$ for fixed sites $i$ and $j$ and there are six bonds. All these couplings transform under some representation $\Gamma_{\rm M}$. We find that $\Gamma_{\rm M}=9 A_g \oplus\ldots$ and these $9$ couplings are further reduced when we consider symmetries that involve translations.


The glide and rotation-translation symmetries reduce the number of linearly independent exchange couplings from $9$ to $6$. They may be divided into three couplings that have $U(1)$ symmetry which are

\begin{align*}
H_{\rm zz} = & J^{(2)}_{\rm zz} \sum_{\langle\langle i,aj,b \rangle\rangle} S^z_{i,a} S^z_{j,b}  \\
H_{\rm xy} = &\frac{J^{(2)}_{\rm xy} }{2} \sum_{\langle\langle i,aj,b  \rangle\rangle}\left( S_{i,a}^+ S_{j,b}^- + S^-_{i,a} S^+_{j,b} \right) \\
H_{\rm DM} = &J^{(2)}_{\rm DM}  \sum_{\langle\langle i,aj,b  \rangle\rangle} \boldsymbol{D}_{a\rightarrow b} \cdot\left( \boldsymbol{S}_{i,a}\times \boldsymbol{S}_{j,b} \right), 
\end{align*}
where $\boldsymbol{D}_{a\rightarrow b}=\pm \hat{\boldsymbol{z}}_c$ along the one-dimensional chains with $+$ for $a=1$ and $b=2$ and minus otherwise. Then there are three exchange operators that are invariant only under discrete symmetries. These are most conveniently represented by considering six particular sites on the puckered hexagon where site $j_1$ is separated from $i$ in the $\hat{\boldsymbol{y}}$ direction:
\begin{align*}
& H_{++} = \frac{J^{(2)}_{++}}{2} \sum_i \left( S_{i,1}^+ S_{j_1 ,2}^+  + e^{-2\pi i /3} S_{i,1}^+ S_{j_2 ,2}^+ + e^{2\pi i /3} S_{i,1}^+ S_{j_3 ,2}^+ + S_{i,1}^+ S_{j_4,2}^+ + e^{-2\pi i /3} S_{i,1}^+ S_{j_5,2}^+ + e^{2\pi i /3} S_{i,1}^+ S_{j_6,2}^+  \right) + {\rm h.c.}  \\
& H_{{\rm z}\pm,{\rm as}} = \frac{J^{(2)}_{{\rm z}\pm,{\rm as}}}{2} \sum_i \left(  S_{i,1}^{+}S_{j_1 ,2}^{z}-S_{i,1}^{z}S_{j_1 ,2}^{+} + \ldots \right) + {\rm h.c.} \\
& H_{{\rm z}\pm,{\rm s}} = \frac{J^{(2)}_{{\rm z}\pm,{\rm s}}}{2} \sum_i  i \left( S_{i,1}^{+}S_{j_1, 2}^{z} + S_{i,1}^{z}S_{j_1, 2}^{+} + \ldots  \right) + {\rm h.c.} 
\end{align*}
While $H_{++}$ is given in full, the ellipses in the expressions for $H_{{\rm z}\pm,{\rm as}}$ and $H_{{\rm z}\pm,{\rm s}} $ indicate those exchange couplings obtained by performing symmetry operations on the one bond joining $i,1$ and $j_1, 2$.

\begin{figure}[htpb]
\includegraphics[width=0.5\columnwidth]{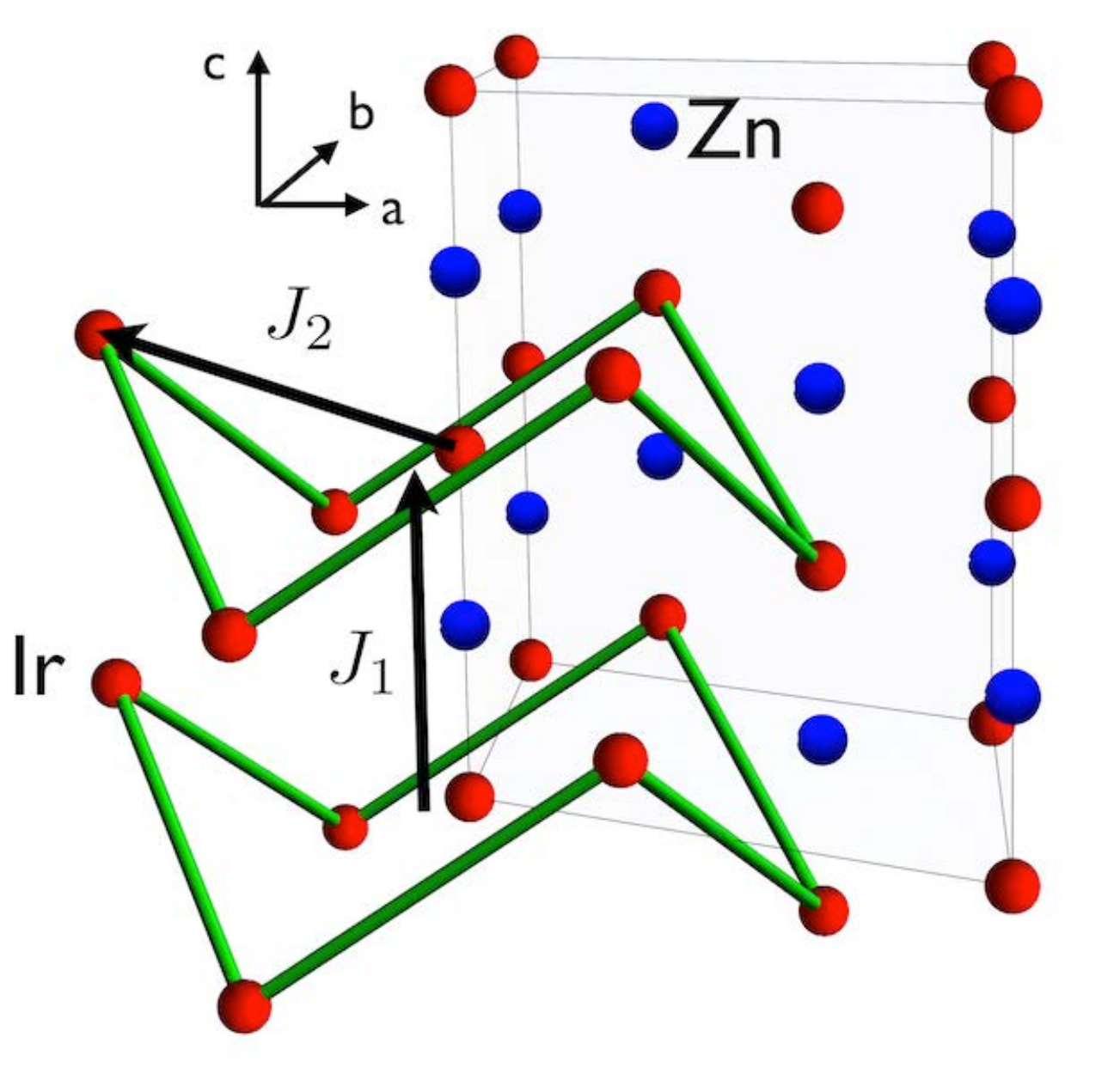}
\caption{ 
{ (Color online) Figure showing the hexagonally coordinated second neighbors about each Ir ion (red spheres). The blue spheres show Zn ions. $J_1$ and $J_2$ shows the magnetic exchanges between Ir ions along the chain and between the chains}.  
\label{fig:Coordination}}
\end{figure}

The correct magnetic structure for \szio\ is obtained as follows. We set $J^{(2)}_{{\rm z}\pm,{\rm s}}=\sqrt{3} J^{(2)}_{{\rm z}\pm,{\rm as}}$ which selects the required structure up to a small rotation of moments in the XY plane and up to a canting angle from the $c$ axis. This constraint is apparently fine-tuned and we conjecture it is fixed by the microscopic superexchange mechanism. The magnetic structure is stable to the presence of Ising couplings to first and second neighbor as shown in the phase diagram Fig.4 in the main paper. The structure is also robust to a nonzero $J^{(2)}_{++}$ coupling and we make use of this fact to fit the spin wave intensity. In order to capture the exact structure we also switch on a small second neighbor $J^{(1)}_{\rm xy}$ coupling. The couplings are given explicitly in the caption to Fig.5 in the main paper. }

%
%
\clearpage
\section*{}
{\bf {\Large {VI. Temperature Dependence of Inelastic Intensity}}}
\begin{figure}[htpb]
\includegraphics[width=0.5\columnwidth, trim={0 0 0 0},clip]{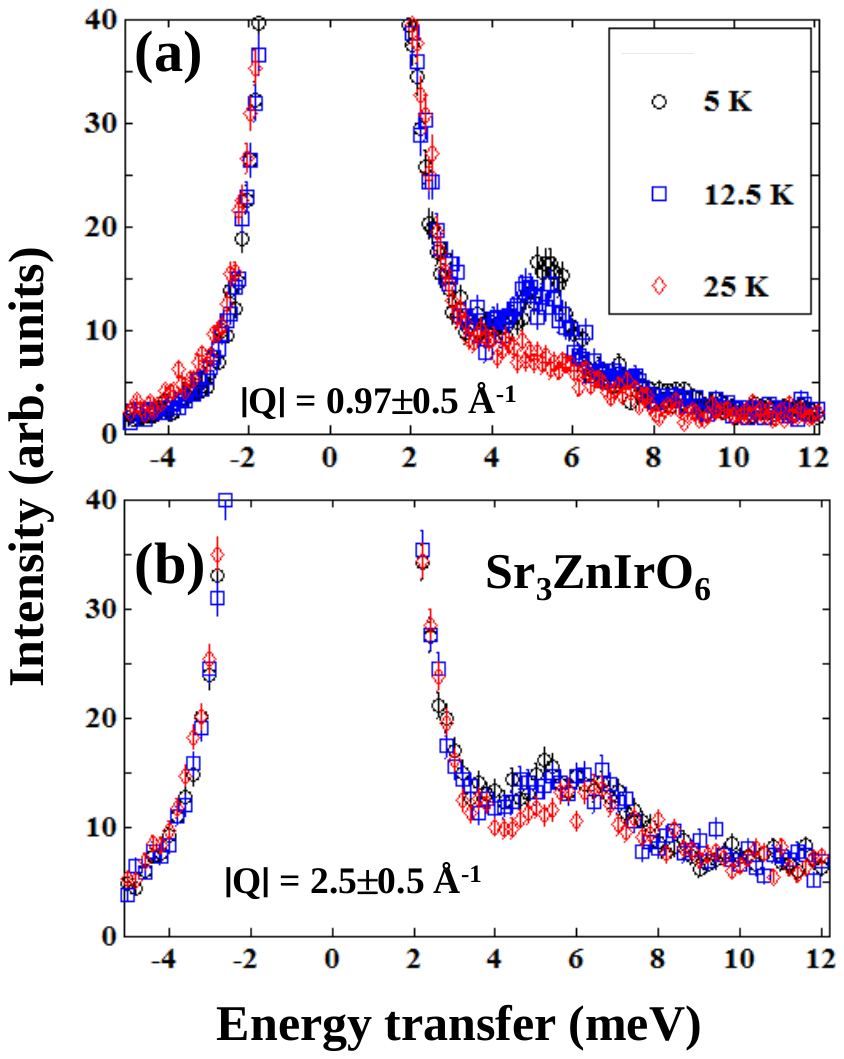}
\caption{ 
{(Color online)} Q-integrated energy cuts at various temperatures (a) Q=0.97$\pm$0.5\AA$^{-1}$ and (b) Q=2.5$\pm$0.5\AA$^{-1}$, which show that the peak near 6 meV at Q=2.5$\pm$0.5\AA$^{-1}$ is non-magnetic as it exists at 25 K.     
\label{fig:cut}}
\end{figure}}


\begin{thebibliography}{99}
\bibitem{SrIrO}
B. J. Kim, H. Jin, S. J. Moon, J.-Y. Kim, B.-G. Park, C. S.
Leem, J. Yu, T.-W. Noh, C. Kim, S.-J. Oh, J.-H. Park, V. Durairaj, G. Cao, and E. Rotenberg, Phys. Rev. Lett. {\bf 101}, 076402 (2008);
Y. K. Kim, N. H. Sung, J. D. Denlinger and B. J. Kim, Nat. Phys. {\bf 12}, 37 (2016);
Y. K. Kim, O. Krupin, J. D. Denlinger, A. Bostwick, E. Rotenberg, Q. Zhao, J. F. Mitchell, J. W. Allen and B. J. Kim,  Science {\bf 345}, 187 (2014).
\bibitem{Jackeli2009} 
G. Jackeli and G. Khaliullin, Phys. Rev. Lett. {\bf 102}, 017205 (2009).
\bibitem{Chaloupka2010} 
J. Chaloupka, G. Jackeli, G. Khaliullin, Phys. Rev. Lett. {\bf 105}, 027204 (2010).
\bibitem{Choi2012}
S. K. Choi, R. Coldea, A. N. Kolmogorov, T. Lancaster, I. I. Mazin, S. J. Blundell, P. G. Radaelli, Yogesh Singh, P. Gegenwart, K. R. Choi, S.-W. Cheong, P. J. Baker, C. Stock, and J. Taylor, Phys. Rev. Lett. {\bf 108}, 127204 (2012).
\bibitem{Chun2015}
S.H. Chun, J-W Kim, J. Kim, H. Zheng, C. C. Stoumpos,
C. D. Malliakas, J. F. Mitchell, K. Mehlawat, Y. Singh, Y. Choi, T. Gog, A. Al-Zein,
M. M. Sala, M. Krisch, J. Chaloupka, G. Jackeli, G. Khaliullin and B. J. Kim
Nat. Phys. {\bf 11}, 463 (2015).
\bibitem{Review}
W. Witczak-Krempa, G. Chen, Y.-B. Kim, and L. Balents, Annu. Rev. Condens. Matter Phys. {\bf 5}, 57 (2014).
\bibitem{GenRef1} See, for instance, S. Rayaprol, K. Sengupta, and E.V. Sampathkumaran, Solid State Commun. {\bf 128}, 79 (2003);  E. V. Sampathkumaran, N. Fujiwara, S. Rayaprol, P.K. Madhu, and Y. Uwatoko, Phys. Rev. B {\bf 70}, 014437 (2004); S. Agrestini, L. C. Chapon, A. Daoud-Aladine, J. Schefer, A. Gukasov, C. Mazzoli, M. R. Lees, and O. A. Petrenko, Phys. Rev. Lett. {\bf 101}, 097207 (2008).
\bibitem{GenRef2} See, for instance, T. Basu, K. K. Iyer, K. Singh, and E. V. Sampathkumaran, Sci. Rep. {\bf 3}, 3104 (2013); A. Jain, P. Y. Portnichenko, H. Jang, G. Jackeli, G. Friemel, A. Ivanov, A. Piovano, S. M. Yusuf, B. Keimer, and D. S. Inosov, Phys. Rev. B {\bf 88}, 224403 (2013).
\bibitem{Lefrancois2014} E. Lefrancois, L. C. Chapon, V. Simonet, P. Lejay, D. Khalyavin, S. Rayaprol, E. V. Sampathkumaran, R. Ballou, and D. T.Adroja, Phys. Rev. B {\bf 90}, 014408 (2014).
\bibitem{Ou2014} X. Ou and H. Wu, Sci. Rep. {\bf 4}, 4609 (2014).
\bibitem{CCO}
H. Kageyama, K. Yoshimura, K. Kosuge, M. Azuma, M. Takano, H. Mitamura and T. Goto, J. Phys. Soc. Jpn {\bf 66}, 1607 (1997).
\bibitem{CCO2}
A. Maignan, V. Hardy, S. Hébert,   M. Drillon, M. R. Lees,  O. Petrenko, D. Mc K. Paul and  D. Khomskii, J. Mater. Chem. {\bf 14}, 1231 (2004).
\bibitem{CCO3}
V. Hardy, M. R. Lees, O. A. Petrenko, D. McK. Paul, D. Flahaut, S. Hébert, and A. Maignan, Phys. Rev. B {\bf 70}, 064424 (2004); V. Hardy, D. Flahaut, M. R. Lees, and O. A. Petrenko, Phys. Rev. B {\bf 70}, 214439 (2004).
\bibitem{CCO4}
C. L. Fleck, M. R. Lees, S. Agrestini, G. J. McIntyre and O. A. Petrenko, Europhysics Letters {\bf 90}, 67006 (2010).
\bibitem{CRO}
S. Niitaka, K. Yoshimura, K. Kosuge, M. Nishi, and K. Kakurai, Phys. Rev. Lett. {\bf 87}, 177202 (2001).
\bibitem{Mikhailova2012} D. Mikhailova, B. Schwarz, A. Senyshyn, A. M. T. Bell, Y. Skourski, H. Ehrenberg, A. A. Tsirlin, S. Agrestini, M. Rotter, P. Reichel, J. M. Chen, Z. Hu, Z. M. Li, Z. F. Li, and L. H. Tjeng, Phys. Rev. B {\bf 86}, 134409 (2012).
\bibitem{Kawasaki}
S. Kawasaki, M. Takano and T. Inam, J. Solid State Chem., {\bf 145}, 302 (1999); S. Rayaprol, K. Sengupta, E. V. Sampathkumaran, and Y. Matsushita, J. Solid State Chem. {\bf 177}, 3270 (2004); 
\bibitem{Choi2008} Y. J. Choi, H. T. Yi, S. Lee, Q. Huang, V. Kiryukhin, and S.-W. Cheong, Phys. Rev. Lett. {\bf 100}, 047601 (2008).
\bibitem{Nguyen1995} T. N. Nguyen and H.-C. zur Loye, J. Solid. Chem., {\bf 117}, 300 (1995).
\bibitem{LampeOnnerud96}
C. Lampe-\"{O}nnerud, M. Sigrist, and H.-C. zur Loye, J. Solid State Chem., {\bf 127}, 25 (1996).
\bibitem{Niazi200x} A. Niazi, E. V. Sampathkumaran, P. L. Paulose, D. Eckert, A. Handstein, and K. H. Mueller, Solid State Commun. {\bf 120}, 11 (2001); A. Niazi, E. V. Sampathkumaran, P. L. Paulose, D. Eckert, A. Handstein, and K.H. Mueller, Phys. Rev. B {\bf 65}, 064118 (2002).


\bibitem{SM} Supplementary Material, a detail on the theoretical model of Sr$_3$ZnIrO$_6$, P. A. McClarty, A. D. Hillier, D. T. Adroja, D. D. Khalyavin, S. Rayaprol, P. Manuel, W. Kockelmann and E. V. Sampathkumaran. 
\bibitem{Hillier2011} A. D. Hillier, D. T. Adroja, W. Kockelmann, L. C. Chapon, S. Rayaprol, P. Manuel, H. Michor, and E. V. Sampathkumaran, Phys. Rev. B. {\bf 83}, 024414 (2011).
\bibitem{Kovalev} O. V. Kovalev (ed.), {\it Representation of the Crystallographic Space Groups} (Gordon and Breach, Newark, NJ, 1993).
\bibitem{FULLPROF} J. Rodriguez-Carvajal, FULLPROF 2k, Version 2.0c, July 2002, Lab. Leon Brillouin, 2002.
\bibitem{Toth2016} S. Toth, W. Wu, D. T. Adroja, S. Rayaprol, and E. V. Sampathkumaran, Phys. Rev. B {\bf 93}, 174422 (2016).
\bibitem{Yamaj2014} Y. Yamaji, Y. Nomura, M. Kurita, R. Arita, and M. Imada
Phys. Rev. Lett. {\bf 113}, 107201, (2014).
\bibitem{DM} I. Dzyaloshinskii, J. Phys. Chem. {\bf 4}, 241 (1958); T. Moriya, Phys. Rev. {\bf 120}, 91 (1960).
\bibitem{Ruan2014} M. Y. Ruan, Z. W. Ouyang, Y. M. Guo, J. J. Cheng, Y. C. Sun, Z. C. Xia, G. H. Rao, S. Okubo and H. Ohta, J. Phys.: Condens. Matter {\bf 26}, 236001 (2014).

\bibitem{Kobayashi2011} K. Kobayashi, T. Nagao, and M. Ito, Acta Cryst., {\bf 67}, 473 (2011).
\bibitem{Unpublished} D. T. Adroja, C. Ritter,  L. Pascut, A. D. Hillier, {\it et al.}, unpublished
\end{thebibliography}
\end{document}